# Prebiotic Fatty Acid Vesicles through Photochemical Dissipative Structuring

Karo Michaelian[1] and Oscar Rodríguez[2]

[1]Department of Nuclear Physics and Applications of Radiation, Instituto de Física, UNAM. Cto. Interior de la Investigación Científica, Ciudad Universitaria, Cuidad de México, C.P. 04510,

karo@fisica.unam.mx, Tel: 01-55-525-6225165

[2]Posgrado en Ciencias Físicas, UNAM. Cto. Interior de la Investigación Científica, Ciudad Universitaria, Cuidad de México, C.P. 04510

nietzsche_31@hotmail.com

**Abstract**

We describe the photochemical dissipative structuring of fatty acids from CO and $CO_2$ saturated water under the solar UVC and UVA photon potential prevalent at Earth's surface during the Archean. Their association into vesicles and their subsequent association with other fundamental molecules of life such as RNA, DNA and carotenoids to form the first protocells is also suggested to occur through photochemical dissipative structuring. In particular, it is postulated that the first vesicles were formed from conjugated linolenic (C18:3n-3) and parinaric (C18:4n-3) acids which would form vesicles stable at the high temperatures (~85 °C) and the somewhat acidic pH values (6.0-6.5) of the Archean ocean surface, resistant to divalent cation salt flocculation, permeable to ions and small charged molecules, but impermeable to short DNA and RNA, and, most importantly, highly dissipative in the prevailing UVC+UVA regions.

**Resumen**

Describimos la estructuración disipativa fotoquímica de los ácidos grasos a partir de agua saturada de CO y $CO_2$ bajo el potencial de los fotones solares en el UVC y UVA prevalentes en la superficie de la Tierra durante el Arcaico. También se sugiere que su asociación en vesículas, y su posterior asociación con otras moléculas fundamentales de la vida, tales como ARN, ADN y carotenoides para formar las primeras protocélulas, se produjeron a través de la estructuración disipativa fotoquímica. En particular, se postula que las primeras vesículas se formaron a partir de ácidos linolenicos conjugados (C18: 3n-3) y parináricos (C18: 4n-3) que darían lugar a vesículas estables a altas temperaturas (~ 85 ° C) y valores algo ácidos de pH (6.0-6.5) en la superficie del océano Archeano, resistentes a la floculación de sal de catión divalente, permeables a iones y pequeñas moléculas cargadas, pero impermeables a ADN y ARN corto, y, lo que es más importante, altamente disipativas en los regiones UVC y UVA predominantes.

Ácidos grasos, vesículas, estructuración disipativa, origen de la vida, protocélulas

## Introduction

Theories concerning the origin and evolution of life must necessarily provide a chemical-physical reason for complexation at all biological levels. The traditional ``survival of the fittest'' paradigm at the macroscopic level or the ``chemical stability'' paradigm at the microscopic molecular level are both deficient when it comes to explaining complexation. A more prudent approach would be to consider only what is well established concerning complexation in chemical-physical systems; that the ordering of material out-of-equilibrium occurs exclusively as a response to the dissipation of an externally imposed generalized thermodynamic potential, or in other words, complexation is concomitant with an increase in global entropy production of the system interacting with its environment. Such ordering of material was given the name ``dissipative structuring'' by Ilya Prigogine [1].

The thermodynamic dissipation theory for the origin and evolution of life [2-5] suggests that every incremental complexation in the history of biological evolution; from the formation of complex prebiotic molecules from simpler precursor molecules, their associations into cellular and multi-cellular organisms, to the hierarchical coupling of biotic with abiotic processes of today's biosphere, must necessarily have coincided with an incremental increase in the dissipation of some externally imposed generalized thermodynamic potential [1,6,7]. This theory identified [2] the important thermodynamic potential driving complexation at life's origin as the long-wavelength UVC solar photon potential arriving at Earth's surface during the Archean (see Fig. 1) since these wavelengths have sufficient free energy to break and reform covalent bonds of carbon-based molecules, but not enough energy to disassociate these molecules [2-4]. The UVA region would also have been important in promoting charge transfer reactions useful in dissipative structuring. The UVC wavelengths reached Earth's surface from before the origin of life and lasting for at least 1,000 million years until the end of the Archean [8,9] (Fig. 1), at which time oxygenic photosynthesis overwhelmed natural oxygen sinks, allowing the formation of an atmospheric ozone layer.

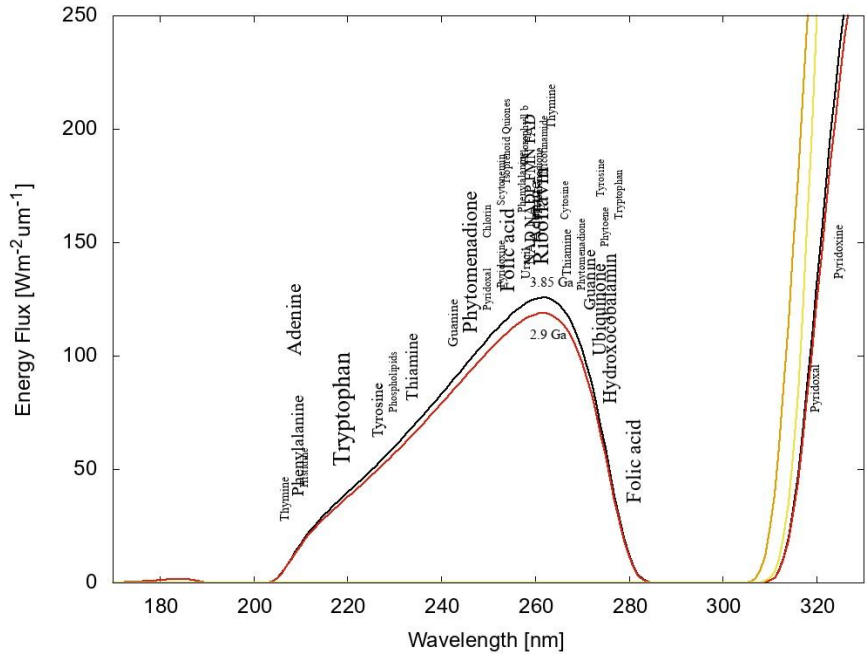

Figure 1. The wavelengths of peaks in the absorption spectra of many of the fundamental molecules of life (common to all three domains) coincide with an atmospheric window (205-285 nm) predicted in the long wavelength end of the UVC region (100-280 nm) at the time of the origin of life approximately 3.85 Ga and until at least 2.9 Ga (curves black and red respectively). Little light was available in the UVB region (280-315 nm) but much in the UVA region (315-400 nm). By around 2.2 Ga (yellow curve), UVC light at Earth's surface had been extinguished by oxygen and ozone resulting from organisms performing oxygenic photosynthesis. The green curve corresponds to the present surface spectrum. Energy flux values correspond to the sun at the zenith. The font size of the letter indicates approximately the relative size of the molar extinction coefficient of the indicated fundamental molecule (UVC pigment). Image credit; adapted with permission from Michaelian and Simeonov [9].

Elsewhere [5], we have analyzed the autocatalytic photochemical production and proliferation of adenine and single strand DNA at the ocean surface under the UVC+UVA Archean photon potential. In this paper, after describing the general chemical and physical characteristics of fatty acids, we analyze the autocatalytic photochemical production and proliferation of fatty acids and their vesicles at the ocean surface under this same photon potential and give non-equilibrium thermodynamic justification for the association of nucleic acids and other fundamental molecules through encapsulation within these first vesicles.

**Characteristics of Fatty Acids**

Natural fatty acids contain a carbon plus hydroxy plus oxygen (carboxyl) head group and a hydrocarbon tail of from 4 to 40 carbon atoms [10] which may be saturated with hydrogen or partially unsaturated (Fig. 2). Fatty acids, fatty alcohols and fatty acid glycerol esters are generally considered to be more relevant to vesicles of early life than the phospholipids composing the cellular wall of most present day organisms because phospholipid biosynthesis is very different between bacteria and archaea, suggesting that their common ancestor was devoid of phospholipid membranes (although this view has been challenged [11,12]). Also, fatty acids are simpler single chain molecules that are more easily produced through abiotic heat activated Fischer-Tropsch or photochemical polymerization of smaller chain hydrocarbons such as ethylene ($C_2H_4$) (Fig. 3). Ethylene itself, and other small chain hydrocarbons, can be derived from the reduction of $CO_2$ or CO employing water as the electron donor and either chemically or photochemically catalyzed. Ethylene can also be produced by the UV photolysis of methane which appears to be an important mode of hydrocarbon production in the atmosphere of Titan [13].

Fatty acids are conserved as components of the cell walls in organisms from all three domains of life. On today's ocean surface, and in the Archean fossil record, there is an abundance of even number carbon fatty acids over odd number and this could be explained if indeed such fatty acids are formed by polymerization (telomerization) of ethylene. In particular, there is a predominance of 16 and 18 carbon atom fatty acids in the whole available Precambrian fossil record [14,15] and, in fact, in today's organisms, and in the aerosols obtained above today's ocean surface [16,17].

In general, the longer the hydrocarbon tail and the greater the degree of saturation, the greater the melting temperature; the degree of saturation usually having the greater influence because double carbon bonds introduce a curvature into the hydrocarbon tail preventing them from bonding strongly with neighbors. For example, lipid desaturation by the enzyme *desaturase* is a key factor in the protection of the photosynthetic apparatus from low-temperature and high-light stresses, as well as being an effective tool for manipulating membrane microviscosity [18]. For example, saturated myristic acid (C14:0) has a melting temperature of 55 °C [19] while linoleic acid with two unsaturated bonds (18:2) has a melting temperature of only -5°C. However, if the unsaturation occurs as conjugated double bonds in the trans configuration near the tail end of the hydrocarbon, such as in parinaric acid (C18:4n-3, see figure 2, then unsaturation is not greatly destabilizing. Parinaric acid has a high melting temperature of 85 °C.

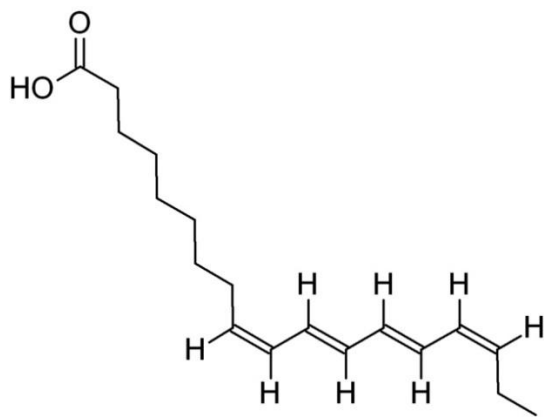

Figure 2. Parinaric acid is a conjugated fatty acid of 18 carbon atoms having a melting temperature of (85-86 °C), roughly the temperature of the Archean ocean surface at the origin of life. The 4 conjugated carbon double bonds lead to delocalized electrons, giving it strong absorption over UVB and UVA region of 310 to 340 nm (Fig. 1). Conjugated hydrocarbons have conical intersections allowing for rapid dissipation to heat of the photon-induced electronic excited singlet state energy, making them excellent candidates for Archean dissipative structures. UVC-induced cross linking with other fatty acids at the site of a double bond would convert this conjugated parinaric acid into conjugated linolenic acid (C18:3n-3) with absorption in the UVC at 269 nm (see Fig. 1).

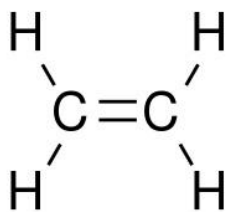

Figure 3. Ethylene. A common precursor molecule for the formation of the hydrocarbon tails of the fatty acids through successive polymerization (telomerization). Ethylene can be derived from the reduction of $CO_2$ or CO employing water as the electron donor, either chemically or photochemically catalyzed. Ethylene can also be produced by the UV photolysis of methane which appears to be an important mode of hydrocarbon production in the atmosphere of Titan.

Direct photon-induced deprotonation under UVC light (254 nm) of fatty acids can lead to conjugation of the carbon bonds of the acyl tail [20]. The steps involved in the conjugation process are i) photon-induced deprotonation giving rise to carbon-carbon double bonds, ii) double bond migration to give a conjugated diene, triene, or tetraene. The greater the degree of unsaturation of the hydrocarbon tail, the greater the probability of forming conjugated bonds under UVC light [20]. The same UVC light, however, can also cause cross linking between adjacent acyl tails at the site of a double bond which would reduce the average conjugation number [21]. Under the constant UVC flux existing at the origin of life, a stationary state of particular conjugation numbers would therefore have arisen. Under the UV light of the Archean, and according to the postulates of the thermodynamic dissipative structuring [5], the conjugation numbers of fatty acids in this stationary state would be either 2, 3, or 4 since these give rise to peak absorption at 233, 269, and 310-340 nm respectively, which all lie within the Archean surface solar spectrum (Fig. 1).

In the following, we will concentrate on vesicles formed from mainly conjugated fatty acids of 18 carbons for the following reasons; 1) fatty acids of 18 (and 16) carbon atoms are predominant in sediments as early as 3.4 Ga and throughout the Archean [14,15] and indeed are the most prevalent sizes in the three domains of today's organisms [15], 2) vesicles containing fatty acids with three conjugated double bonds required for light absorption around the peak in the Archean UVC spectrum at 260 nm (see below) must contain fatty acids of at least 18 carbon atoms in order to be stable at high surface temperatures of the early Archean of around 85 °C, and 3) apart from the telomerization of ethylene, there is a rather simple photochemical route to the production of 18 carbon atom fatty acids from the UVC induced polymerization of highly surface active 9 carbon nonanoic acid $C_9H_{18}O_2$ which can be produced through UVC photochemical dissipative structuring of CO, $CO_2$ saturated water [22]. Our postulate for the fatty acids of the first protocells, with chain lengths of 18 carbon atoms and unsaturated and conjugated, is distinct to that previously asserted as being the most plausible prebiotic fatty acids for the first protocells -- of saturated short chain (≤ 10 carbon atoms) lengths (such as myristoleic acid and its alcohol) because these are the most likely to arise from a Fischer-Tropsch type chemistry [23]. This conclusion, however, ignores UVC induced surface chemistry for synthesis, polymerization, and cross linking (see below) of fatty acids, and the fact that a greater thermal stability of the vesicles then generally believed would have been required because of the high surface temperatures of the early Archean. Inclusion of even longer chain lengths (≥ 20 carbon atoms, for which there is some evidence in the early fossil record (see [14,24] and references therein) increases still the stability with respect to pH range [25] and temperature.

**Photochemical Synthesis of Fatty Acids**

A plausible alternative to ocean floor Fischer-Tropsch synthesis of the hydrocarbon chains is ocean surface photochemical synthesis. The fatty acid hydrocarbon tails can be built up from the sequential photon-induced polymerization of an initiator molecule such as ethylene (Fig. 3). Photopolymerization occurs through photon-induced sensitization in the UV region of the spectrum. It generally occurs through direct photon-induced cleavage of the initiator molecule producing a free-radical which subsequently attacks the carbon-carbon double bonds of an existing polymer, thus initiating further polymerization [26]. Polymerization rates are more than two orders of magnitude larger at wavelengths of 254 nm (UVC) than at 365 nm (UVA) [26]. Oxygen acts as a strong inhibitor to polymerization by rapidly reacting with the radical to form a peroxy-based radical which does not promote polymerization [26]. Such an oxidation reaction following hydrolysis is the origin of the carboxyl head group of the fatty acids. The presence of oxygen and the lack of surface UVC light today (Fig. 1) means that hydrocarbon chain polymerization at today's ocean surface [16,17] is only a mere ghost of what it probably was at the origin of life.

Indications that ultraviolet light may have played an important role in the formation of hydrocarbons have come from different experiments. Since the early 1960's it was known that irradiation with UVC light of $CO_2$ saturated water containing ferrous salts results in the production of formic acid and formaldehyde [27]. Later, $C_1$ hydrocarbons such as methane, methanol, ethanol and formaldehyde, and formic acid were produced from $CO_2$ and $H_2O$ in a photoelectrochemical reactor consisting of a $TiO_2$-coated electrode suspended in $CO_2$ saturated aqueous solution and subjected to UVC light [28,29]. Klein and Pilpel (1973) [30] demonstrated that short chain amphiphiles can be synthesized by a light-dependent reaction from common simple hydrocarbons and an aqueous film of poly aromatic hydrocarbons (PAHs), ubiquous throughout the universe [9], acting as photosensitizers. It has also been shown that imidazole or porphoryins (pyridine) under UVC light can act as a catalyst for the reduction of $CO_2$ to n=2 and longer chain hydrocarbons [31].

Later, Varghese et al. [32] and also Roy et al. [33] identified longer chain hydrocarbons in similar mixtures of $CO_2$ and water irradiated with UV light. In general, and in contrast to the Fischer-Tropsch process, the higher the temperature, the higher the pressure, or the greater the ratio of $CO_2$ to $H_2O$, the larger the quantity of longer chain hydrocarbons obtained.

Another plausible route to the formation of fatty acids under a UV environment has been observed Botta et al. [33]. By impinging UV and visible light (185 to 2000 nm) from an xenon lamp on formaldehyde (a product of UVC light on HCN and water) using ZnO and $TiO_2$ as photocatalysts, Botta et al. found that this resulted in fatty acids of chains of from 2 to 5 carbon atoms as well as a host of other fundamental molecules of life such as nucleic acids, and amino acids, as well as glycolaldehyde, a precursor to sugars through the formose-reaction. The mechanism of formation of these fatty acids suggested by Botta et al. follows that proposed earlier by Eschenmoser [34] entailing, first the generation of HCN from formaldehyde followed by its oligomerization to diaminomaleonitrile DAMN (detected in their reaction mixture), hydrolysis and successive electron transfer processes. The formation of DAMN from HCN requires the absorption of photons in the long wavelength UVC region and is exactly the same process which we have described as the photochemical dissipative structuring of the purines from UVC light on HCN and water [5] first observed by Ferris and Orgel [35]. The ZnO and $TiO_2$ photocatalysts are suspected of providing electron transfer reactions after absorbing a UV photon. Although these catalysts are rather common and occur naturally and would have been available on the ocean surface (but in a much more diluted phase than that employed by Botta et al.) another electron donor which could have equally catalyzed the reaction is the amino acid histidine or its intermediate known as amino-imidazole-carbon-nitrile (AICN) and those amino acids with a negatively charged R-group; Glu monosodium salt (Glu•Na), and Asp potassium salt (Asp•K). These, we have argued [36], were among the first amino acids to have formed complexes with RNA and DNA in order to increase global UVC dissipation as evidenced by their charge transfer absorption spectrum peaking at 270 nm [37] falling in the middle of the Archean UVC atmospheric window, and as evidenced by their chemical affinity their codons and/or anticodons [38].

Here we propose that the long chain, n ≥18, hydrocarbons which became incorporated into the fatty acid vesicles of early life were produced by such surface-sensitized UVC-induced polymerization processes acting on shorter chain hydrocarbons such as ethylene produced through UVC light on $CO_2$ saturated surface water or formaldehyde and HCN saturated water at high temperature (> 85 °C) and possibly a higher than present atmospheric pressure (up to 2 bar [39]). In contradistinction to the Fischer-Tropsch polymerizations operating at very high temperatures and pressures (perhaps occurring at deep sea hydrothermal vents) such photochemical polymerization could only have occurred on the ocean surface where concentrations of fatty acids would have been sufficiently high enough to allow access to the reactive triplet state, and where UVC light was most intense. The experiments of Rossignol et al. [22] suggest that catalysts are not needed for such surface polymerizations, however, the existence of catalyst transition metals for the reduction of $CO_2$ such as Fe, Mn, Co, Ni, Cu, or Zn (particularly Fe) would have been available at the Archean ocean surface [40] and these would have undoubtedly increased the rates of hydrocarbon chain growth. There is again also the possibility that an imidazole, such as the amino acid histidine or its intermediate AICN formed in the process of UVC microscopic dissipative structuring of the purines from HCN [5], could have been the reduction agent. It is well known that imidazole is a strong catalyst that can act as either an acid or a base and furthermore, direct evidence has been obtained for its catalytic function in the formation of phosphatic acids from simpler compounds [41].

**Dissipative Structuring**

Saturated fatty acids do not absorb in the UV except for disassociation at < 180 nm and the carboxyl head group which absorbs with a peak at 207 nm [42]. Under the Archean UVC flux photon-induced deprotonation could lead to a double carbon bond forming at any point on the hydrocarbon tail. A single double carbon bond in the tail will absorb at 210 nm. Migration of the double bonds along the tail is known to occur, leading to conjugated bonds [20]. Two double bonds in a conjugated configuration (diene) will lead to strong absorption at 233 nm [20], those having three in a conjugated configuration (triene) will lead to absorption at 269 nm, while those with 4 (tetraene, see Fig. 2) will lead to absorption at 310-340 nm. All of these latter three absorptions lie within the important UVC+ UVA spectrum arriving at Earth's surface during the Archean (see Fig. 1).

Hydrocarbons having conjugated dienes, trienes, or tetraenes almost always have conical intersections [43] allowing rapid dissipation of the electronic excited singlet state energy. Reaching the conical intersection when in the electronic excited state involves a twisting about two C=C bonds and decreasing one of the C-C-C angles producing a kink in the carbon backbone [43].

Therefore, the region of the Archean surface spectrum that structured the CO, $CO_2$ and $H_2O$ into fatty acids are the same photons that will be dissipated efficiently by the final photochemical product. These photochemically synthesized and conjugated fatty acid structures can thus be identified as microscopic dissipative structures [5]. The steps involved in the dissipative structuring are thus the following i) UVC-induced reduction of $CO_2$ and CO in water saturated with these to form ethylene, ii) consecutive UVC-induced telomerization of ethylene to form long hydrocarbon tails of an even number of carbon atoms, iii) oxidation and hydrolysis events to stop the growing of the chain and form the carboxyl group, iv) UVC-induced deprotonation of the tails to form a double bond, v) double bond migration to give a conjugated diene, triene, or tetraene which has its own conical intersection (Fig. 4). These dissipatively structured fatty acids are, of course, robust to further photochemical reactions because of the sub picosecond decay times of their electronic excited states, due to their conical intersections. These decay times are too fast to allow appreciable further chemical transformation.

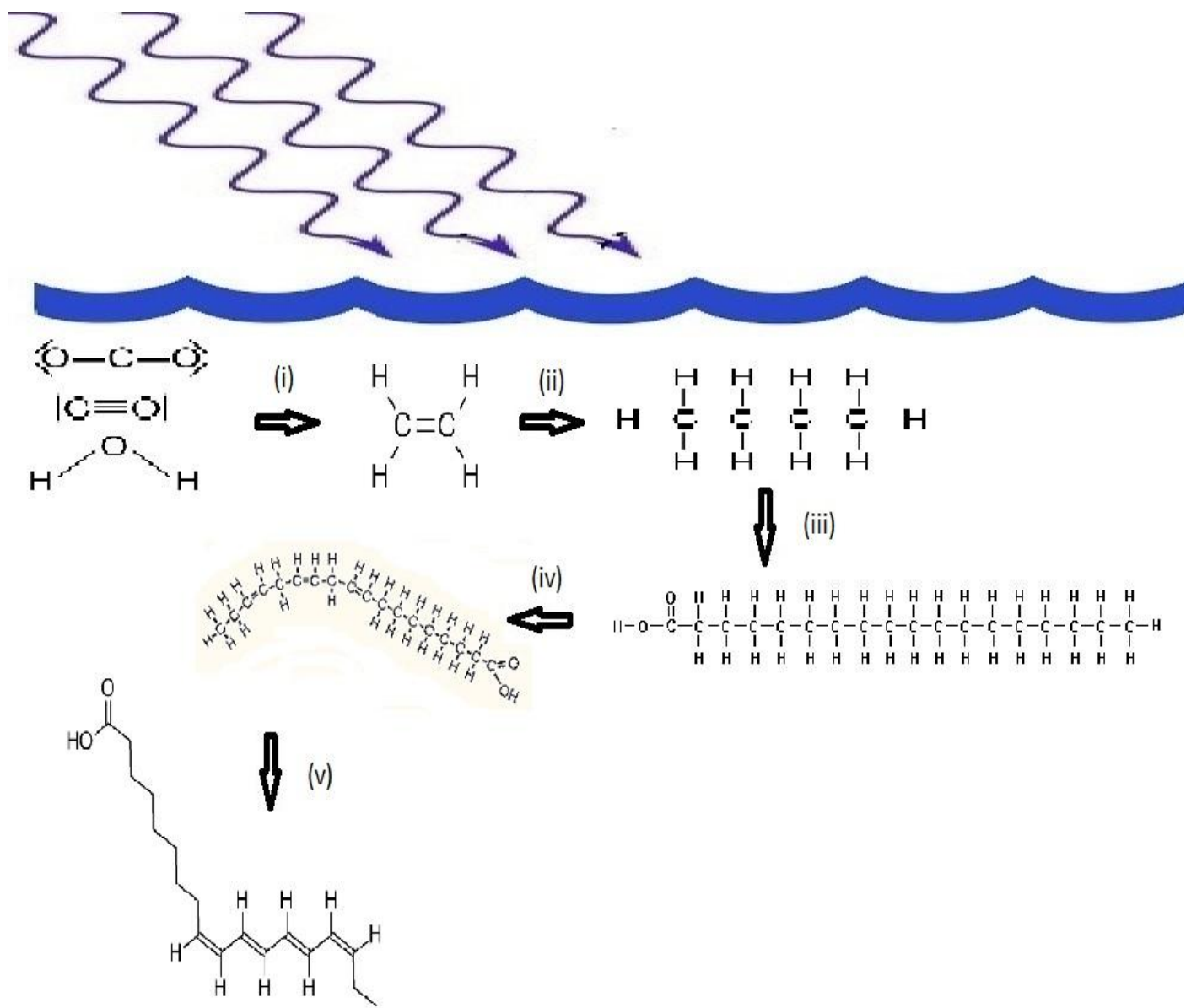

Figure 4. Five steps are involved in the photochemical dissipative structuring of α-parinaric acid under UVC+UVA light, starting from water saturated with $CO_2$ or CO. i) UVC-induced reduction of $CO_2$ and CO in water saturated with these to form ethylene, ii) UVC-induced telomerization of ethylene to form long hydrocarbon tails of an even number of carbon atoms, iii) oxidation and hydrolysis events to stop the growing of the chain and form the carboxyl head group, iv) UVC- or UVA-induced deprotonation of the tails to form double bonds, v) double bond migration to give a conjugated tetraene α-parinaric acid (C18:4n-3) with a conical intersection.

We have identified a similar dissipative structuring route for the synthesis of the nucleobase adenine from HCN under the same UVC photon potential [5]. In this case, the steps involve i) a thermal exothermic process leading from 4HCN molecules to the stable HCN tetramer cis-DAMN, ii) a UVC-induced cis to trans transformation through rotation around a carbon-carbon double bond, iii) a UV-A induced tautomerization, and iv) a UVC-induced ring closure. We believe that similar dissipative structuring routes involving different UVC- or UVB-induced processes exist for all of the fundamental molecules of life and this explains their strong absorption in the UVC (Fig. 1), and, in fact, gives a chemical-physical explanation for the origin of life based on non-equilibrium thermodynamics imperatives.

## Vesicle Formation

The most notable problems associated with the proposal of fatty acids conforming walls of the first protocell are that; 1) vesicles of these form only in a narrow alkaline pH range, 2) they have a tendency to aggregate and crystallize in high salt conditions (salt flocculation), leading some to conclude that life must have started in a fresh water environment [44], and 3) a critical vesiculation concentration (CVC) of the fatty acids is required for spontaneous formation of vesicles and this concentration is considerably higher than that required for phospholipid vesicle formation.

Recent experiments have shown, however, that the pH range of stability and resistance to salt flocculation of fatty acid vesicles can be greatly increased through covalent cross linking among neighboring chains, which can be induced by either UVC light [21], or by moderate temperatures (~50 – 70 °C), or by simple aging [21,45]. The same heat treatment applied to non-conjugated unsaturated fatty acids does not appear to have an as large effect on the increase in range of pH stability or stability against salt flocculation. Under all conditions, vesicles of heat treated conjugated linoleic acid showed better stability than non-conjugated linoleic acid [45]. At the high surface temperatures and large UVC flux of the early Archean, considerable cross-linking between fatty acids could be expected, and conjugated fatty acids would thus form robust vesicles under the high temperatures, low pH, and high salt conditions of the Archean ocean surface. Furthermore, mixtures of fatty acids with fatty alcohols and fatty acid glycerol esters of differing lengths provide surprising resistance to salt flocculation [23]. Esterification (replacing the hydrogen of the OH in the carboxyl head group with, for example methyl $CH_3$) of fatty acids which helps with resistance to salt flocculation (see section "Characteristics of Fatty Acids") can be induced by UVC activated transition metal catalysts to provide electrons for the reduction of the fatty acid [46]. UV radiation has also been shown to lead to the formation of aldehydes from fatty acids (through removal of the oxygen in the OH) [22].

There is also another more likely possibility giving rise to vesicle stability over a greater range of pH. Rather than the end to end model of the hydrophobic tails of phospholipids, the observed structure for fatty acids appears to be more like side-by-side [47] with overlap of respective double bonds (see figure 5). This would allow for a natural aging, or temperature- or UVC induced-, cross linking among fatty acids, which makes these vesicles stable over a very wide range of pH values, from 2 to 14 [47].

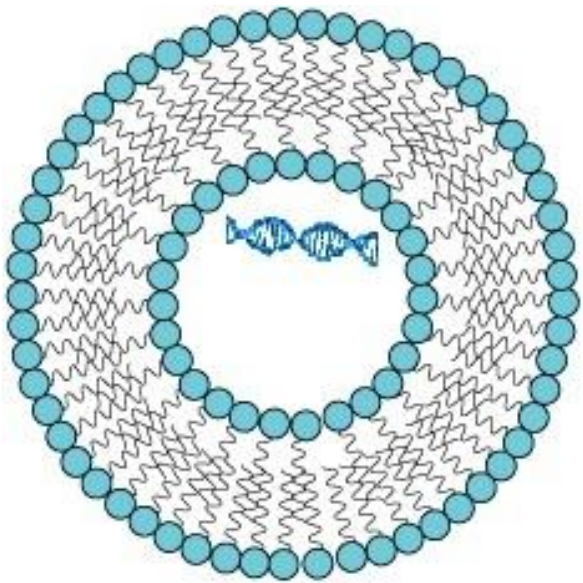

Figure 5. A vesicle made from variously conjugated (n≥18) fatty acids, for example, α-parinaric (C18:4n-3) and conjugated linolenic acid (C18:3n-3), with an encapsulated DNA oligo. The DNA acts as an acceptor molecule for resonant energy transfer from the electronically excited un-conjugated fatty acids which do not have their own conical intersection. The inner and outer layers of the hydrocarbon tails have a side-by-side (rather than end-to-end) orientation which facilitates UVC-induced cross-linking at the overlaps, giving greater stability to the vesicle at the somewhat acid pH values of the Archean ocean surface. The whole system is an efficient photon dissipating system in the UVC+UVA regions of the Archean surface solar spectrum (Fig. 1).

**Conclusions**

We have described a plausible photochemical route to the dissipative structuring of the earliest Archean long chain fatty acids and their vesicles from $CO_2$ and CO saturated water under the long wavelength UVC light prevailing at Earth's surface at the origin of life and throughout the Archean. Ethylene would be a common product synthesized by this UVC light incident on the surface water and these would polymerize (telomerize) under the same light into longer chain hydrocarbons of mainly an even number of carbon atoms. These would stop growing after an oxidation or hydrolysis event. They would then become conjugated through the dissipative structuring under the UVC light giving rise to conjugated parinaric acid (C18:4n-3) and linolenic acid (C18:3n-3). Because of the high surface temperatures (~85 °C), only long chain (≥ 18 carbon atoms) with conjugation near the end of the tail (n-3) would bind sufficiently together to form stable vesicles. This is in agreement with the abundance of this size in the fossil record of fatty acids found in sediments of the early Archean. Cross-linking would occur at the overlap of the inner and outer wall hydrocarbon tails at the site of a double bond, making the vesicles stable over low pH ranges (6.0-6.5) and high salt conditions of the Archean ocean surface.

Many of the fatty acids would have their own conical intersection allowing the rapid dissipation of the photon-induced electronic excitation energy. These vesicles would then be performing the non-equilibrium thermodynamic function of dissipating into heat the prevailing UVC+UVA photon flux reaching Earth's surface. This was the initial driving force for complexation of material, from synthesis, to proliferation and evolution; all through dissipative structuring. For example, chemical affinity of the fatty acids, mediated through divalent cations, with RNA and DNA, or with the carotenoids, would provide for greater dissipation, especially for those non-conjugated fatty acids which did not have their own conical intersections. Resonant energy transfer would allow the electronic excitation energy of the fatty acids to be dissipated through the conical intersection of the nucleic acid or the carotenoids. This scenario provides a physical-chemical foundation based on photon dissipation for the origin, proliferation and evolution of the fatty acids in conjunction with other fundamental molecules of life.

## Acknowledgments

The authors are grateful for comments on the manuscript by C. Bunge and A. Simeonov. The financial support of DGAPA-UNAM project number IN102316 is greatly appreciated. O.R. further acknowledges DGAPA and CONACyT for providing scholarships during various stages of this work.

## Bibliography

[1] Prigogine, I. *Introduction to Thermodynamics Of Irreversible Processes*. 3rd Edition, New York: John Wiley & Sons, 1967.
[2] Michaelian, K.. "Thermodynamic origin of life", *ArXiv* 2009 (http://arxiv.org/abs/0907.0042)
[3] Michaelian, K. "Thermodynamic dissipation theory for the origin of life", *Earth Syst. Dynam.* 2011, **224**, 37-51.
[4] Michaelian, K. *Thermodynamic Dissipation Theory of the Origina and Evolution of Life: Salient characteristics of RNA and DNA and other fundamental molecules suggest an origin of life driven by UV-C light*. Mexico City: Self-published. Printed by CreateSpace, 2016, ISBN:9781541317482.
[5] Michaelian, K. "Microscopic dissipative structuring and proliferation at the origin of life", *Heliyon*. 2017, **3**, e00424.
[6] Onsager, L. "Reciprocal Relations in Irreversible Processes, I", *Phys. Rev.* 1931, **37**, 405-426.
[7] Onsager, L. "Reciprocal Relations in Irreversible Processes, II", *Phys. Rev.* 1931, **38**, 2265.

[8] Sagan, C. "Ultraviolet Selection Pressure on the Earliest Organisms", *J. Theor. Biol.* 1973, **39**, 195-200.

[9] Michaelian, K. and Simeonov, A. "Fundamental molecules of life are pigments which arose and co-evolved as a response to the thermodynamic imperative of dissipating the prevailing solar spectrum", *Biogeosciences*. 2015, **12**, 4913-4937.

[10] Johnson, D.W. "A synthesis of unsaturated very long chain fatty acids", *Chemistry and Physics of Lipids*. 1990, **56** (1), 65-71. ISSN: 0009-3084

[11] Pereto, J., Lopez-Garcia, P. and Moreira, D. "Ancestral lipid biosynthesis and early membrane evolution", *Trends Biochem. Sci.* 2004, **29**, 469-477.

[12] Lombard, J., López-García, P. and Moreira, D. "The early evolution of lipid membranes and the three domains of life", *Nature Reviews, Microbiology.* 2012, **10**, 507-515.

[13] Bar-nun, A. and Podolak, M. "The photochemistry of hydrocarbons in titan's atmosphere", *Icarus*. 1979, **38** (1), 115-122.

[14] Han, J. and Calvin, M. "Occurrence of fatty acids and aliphatic hydrocarbons in a 3.4 billion-year-old sediment", *Nature*. 1969, **224** (5219), 576-577.

[15] Van Hoeven, W., Maxwell, J.R. and Calvin, M. "Fatty acids and hydrocarbons as evidence of life processes in ancient sediments and crude oils", *Geochimica et Cosmochimica Acta*. 1969, **33** (7), 877-881.

[16] Mochida, M., Kitamori, Y., Kawamura, K., Nojiri, Y. and Suzuki, K. "Fatty acids in the marine atmosphere: Factors governing their concentrations and evaluation of organic films on sea-salt particles", *Journal of Geophysical Research: Atmospheres*. 2002, **107** (D17), AAC 1-1-AAC 1-10.

[17] Wellen, B. A., Lach, E. A. and Allen H. C. "Surface pka of octanoic, nonanoic, and decanoic fatty acids at the air-water interface: applications to atmospheric aerosol chemistry", *Phys. Chem. Chem. Phys.* 2017, **19**, 26551-26558.

[18] Kis M, Zsiros O, Farkas T, Wada H, Nagy F, Gombos Z. "Light-induced expression of fatty acid desaturase genes", *Proc Natl Acad Sci USA* 1998, **95**, 4209-4214.

[19] Budin, I., Prywes, N., Zhang, N. and Szostak J. W. "Chain-length heterogeneity allows for the assembly of fatty acid vesicles in dilute solutions", *Biophysical Journal*. 2014, **107**, 1582-1590.

[20] Mandal, T. K. and Chatterjee, S. N. "Ultraviolet- and sunlight-induced lipid peroxidation in liposomal membrane", *Radiation Research*. 1980, **83**, 290-302.

[21] Bassas, M., Marqués, A. M. and Manresa, "A. Study of the crosslinking reaction (natural and uv induced) in polyunsaturated pha from linseed oil", Biochemical Engineering Journal 2007 40:275{283, 2007.

[22] Rossignol, S., Tinel, L., Bianco, A.and Passanant, Mi. "Atmospheric photochemistry at a fatty acid-coated air-water interface", *Science*. 2016, **353**, 699-702.

[23] Mansy S. S. and Szostak J. W. "Thermostability of model protocell membranes", *PNAS*. 2008, **105** (36), 13351-13355.

[24] Hahn, J. and Haug, P. "Traces of Archaebacteria in Ancient Sediments", *System. Appl. Microbiol.* 1986, **7,** 178-183.

[25] Aveyard, R., Binks, . B.P., Carr, N. and Cross A.W. "Stability of insoluble monolayers and ionization of langmuirblodgett multilayers of octadecanoic acid", *Thin Solid Films*. 1990, **188** (2), 361-373.

[26] Bowman, C. N. and Kloxin, C. J. "Toward an enhanced understanding and implementation of photopolymerization reactions", *AIChE J.* 2008, **54**, 2775-2795.

[27] Getoff, N. "Reduktion der kohlensäure in wässeriger lösung unter einwirkung von uv-licht", *Zeitschrift für Naturforschung B*. 1962, **17** (2), 87-90.

[28] Halmann, M. "Photoelectrochemical reduction of aqueous carbon dioxide on p-type gallium phosphide in liquid junction solar cells", *Nature*. 1978, **275**, 115-116.

[29] ULMAN, M., TINNEMANS, A. H. A., MACKOR, A., AURIAN-BLAJENI, B. andHALMANN M." Photoreduction of carbon dioxide to formic acid, formaldehyde, methanol, acetaldehyde and ethanol using aqueous suspensions of strontium titanate with transition metal additives", *International Journal of Solar Energy*. 1982, **1** (3), 213-222.

[30] Klein, A. E. and Pilpel, N. "Oxidation of n-alkanes photosensitized by 1-naphthol", *J. Chem. Soc Faraday Trans. 1* 1973, **69**, 1729-1736.

[31] Bocarsly, A. B., Gibson, Q.D., Morris, A. J., L'Esperance, R. P., Detweiler, Z. M., Lakkaraju, P. S., Zeitler, E. L. and Shaw T. W. "Comparative study of imidazole and pyridine catalyzed reduction of carbon dioxide at illuminated iron pyrite electrodes", *ACS Catalysis*. 2012, **2**, 1684-1692.

[32] Varghese, O. K.,  Paulose, M., LaTampa., T. J., Grimes, C. A. "High-Rate Solar Photocatalytic Conversion of CO2 and Water Vapor to Hydrocarbon Fuels", *Nano Letters* 2009, **9** (2), 731-737.

[33] Botta, L., Bizzarri, B.M.,  Piccinino, D., Fornaro, T., Brucato, J.R. and Saladino, R.. "Prebiotic synthesis of carboxylic acids, amino acids and nucleic acid bases from formamide under photochemical conditions", *Eur. Phys. J. Plus*. 2017, **132**, 317.

[34] Eschenmoser, A. "On a Hypothetical Generational Relationship between HCN and Constituents of the Reductive Citric Acid Cycle", *Chemistry & Biodiversity*. 2007, **4** (4), 554-573.

[35] Ferris, J. P. and Orgel, L. E. An "Unusual Photochemical Rearrangement in the Synthesis of Adenine from Hydrogen Cyanide", *J. Am. Chem. Soc.*, 1966, **88**, 1074.

[36] Mejía, J. and Michaelian, K. "Information encoding in nucleic acids through a dissipation-replication relation", ArXiv:1804.05939 [physics.bio-ph] 2018.

[37] Prasad, S., Mandal, I., Singh, S., Paul, A., Mandal, B., Venkatramani, R. and Swaminathan, R.. "Near uv-visible electronic absorption originating from charged amino acids in a monomeric protein", *Chem. Sci*. 2017, **8**, 5416-5433.

[38] Yarus, M., Widmann, J. and Knight, R. "RNA-Amino Acid Binding: A Stereochemecal Era for the Genetic Code", *J Mol Evol*. 2009, **69**, 406-429. DOI 10.1007/s00239-009-9270-1

[39] Som, S. M., Catling, D. C., Harnmeijer, J. P., Polivka, P. M. and Buick, R.. "Air density 2.7 billion years ago limited to less than twice modern levels by fossil raindrop imprints", *Nature*. 2012, **484**, 359-362.

[40] Hardy. J. T. "The sea-surface microlayer (1982) biology, chemistry and anthropogenic enrichment", *Prog. Oceanogr*. 1982, **11,** 307-328.

[41] Epps, D. E. Sherwood, E., Eichberg, J. and Oró, J. "Cyanamide mediated syntheses under plausible primitive earth conditions", *Journal of Molecular Evolution*. 1978, **11**, 279-292.

[42] Vicente, A., Antunes, R., Almeida, D., Franco, I. J. A., Hoffmann, S. V., Mason, N. J., Eden, S., Duflot, D., Canneaux, S., Delwiche, J., Hubin-Franskin, M.-J., and Limão-Vieira, P. "Photoabsorption measurements and theoretical calculations of the electronic state spectroscopy of propionic, butyric, and valeric acids", *Phys. Chem. Chem. Phys.*2009,**11** (27),5729-5741.

[43] Celani, P., Garavelli, M., Ottani, S., Bemardi, F., Robb, M. A. and Olivucci, M. "Molecular "Trigger" for Radiationless Deactivation of Photoexcited Conjugated Hydrocarbons", *J. Am. Chem. Soc.* 1995, **117**, 11584-11585

[44] Milshteyn, D., Damer, B., Havig, J. and Deamer, D.. "Amphiphilic compounds assemble into membranous vesicles in hydrothermal hot spring water but not in seawater", *Life*, 2018, **8**,11.

[45] Fan, Y., Fang, Y., Ma L., and Jiang, H. "Investigation of micellization and vesiculation of conjugated linoleic acid by means of self-assembling and self-crosslinking", *J. Surfact. Deterg.* 2015, **18**, 179-188.

[46] Verma P., Kaur, K., Wanchoo, R. K., Toor, A. P. "Esterification of acetic acid to methyl acetate using activated TiO2 under UV light irradiation at ambient temperature", *Journal of Photochemistry and Photobiology A: Chemistry.* 2017, **336**, 170-175.

[47] Fan, Y. Fang, Y. and Ma L. "The self-crosslinked ufasome of conjugated linoleic acid: Investigationof morphology, bilayer membrane and stability", *Colloids and Surfaces B: Biointerfaces* 2014, **123**, 8-14.